\documentclass[preprint]{aastex}

\usepackage{amsmath,amssymb,bm}

\DeclareFontFamily{OMS}{bcmsy}{}
\DeclareFontShape{OMS}{bcmsy}{m}{n}{%
	<5> bcmsy5
	<6> bcmsy6
	<7> bcmsy7
	<8> bcmsy8
	<9> bcmsy9
	<10-> bcmsy10}{}
\DeclareSymbolFont{symbols}{OMS}{bcmsy}{m}{n}

\DeclareFixedFont{\elevensy}{OMS}{bcmsy}{m}{n}{10.95pt}
\DeclareFixedFont{\ninesy}{OMS}{bcmsy}{m}{n}{9pt}

\newcommand{\<}{{\kern-5pt}}

\newcommand{\thrj}[6]{\biggl(
	\arraycolsep .2em
	\begin{matrix}
	#1&#2&#3\\
	#4&#5&#6\\
	\end{matrix}\biggr)}
\newcommand{\sixj}[6]{\biggl\{
	\arraycolsep .2em
	\begin{matrix}
	#1&#2&#3\\
	#4&#5&#6\\
	\end{matrix}\biggr\}}

\def\mlangle{\kern.175em\langle}	
\def\mrangle{\rangle\kern.175em}	
\def\ket#1{|{#1}\mrangle}
\def\bra#1{\mlangle{#1}|}
\def\redmat#1#2#3{\bra{#1}\kern -1pt|#2|\kern -1pt\ket{#3}}

\newcommand{\apx}[1]{^{\mbox{\tiny{(#1)}}}}

\begin{document}

\title{Rayleigh Scattering in Spectral Series with $L$-Term Interference}

\author{R.\ Casini$^a$, R.\ Manso Sainz$^b$, and T.\ del Pino
Alem\'an$^a$}

\affil{$^a$High Altitude Observatory, National Center for Atmospheric
Research,\footnote{The National Center for Atmospheric Research is sponsored
by the National Science Foundation.}\break
P.O.~Box 3000, Boulder, CO 80307-3000, U.S.A.}
\affil{$^b$Max-Planck-Institut f\"ur Sonnensystemforschung,\break
Justus-von-Liebig-Weg 3, 37077 G\"ottingen, Germany}

\begin{abstract}
We derive a formalism to describe the scattering of polarized radiation 
over the full spectral range encompassed by atomic transitions belonging 
to the same spectral series (e.g., the \ion{H}{1} Lyman and Balmer series, 
the UV multiplets of \ion{Fe}{1} and \ion{Fe}{2}). This allows us to 
study the role of radiation-induced coherence among the upper terms of 
the spectral series, and its contribution to Rayleigh scattering and the
polarization of the solar continuum. We rely on previous theoretical results 
for the emissivity of a three-term atom of the $\Lambda$-type taking into 
account partially coherent 
scattering, and generalize its expression in order to describe a
``multiple $\Lambda$'' atomic system underlying the formation of a spectral series. 
Our study shows that important polarization
effects must be expected because of the combined action of partial
frequency redistribution and radiation-induced coherence among 
the terms of the series. In particular, our model predicts the correct 
asymptotic limit of 100\% polarization in the far wings of a 
\emph{complete} (i.e., $\Delta L=0,\pm 1$) group of transitions, which 
must be expected on the basis of the principle of spectroscopic stability.
\end{abstract}

\section{Introduction}

The observation and analysis of the polarization signatures of resonant 
transitions in the solar spectrum have proven to be extremely useful for 
developing diagnostic tools of the magnetism of the top layers of the
solar atmosphere (upper photosphere, chromosphere, transition region;
see, e.g., the reviews by \citealt{St15,TB17}). 
%
Relatively recent discoveries that have been fostered by adding polarization
information to the spectroscopic analysis of the solar radiation---such as the
so-called ``second solar spectrum'' \citep{Iv91,SK97}---have revealed the potential
of these diagnostics. Often, the observed polarization patterns have defied 
our ability to adequately model them, at times giving rise to
``enigmas'' about how the magnetism of the solar atmosphere is realized, and
even to the point of questioning the very foundations of our theoretical
understanding of polarized line formation \citep{St11}.

The modeling of the scattering polarization in the solar spectrum becomes 
particularly challenging in the UV, because of the high density of 
spectral features observed there. 
On the other hand, Rayleigh scattering in stellar atmospheres is 
dominated by contributions in the UV, particularly in the wings of
\ion{H}{1} Ly$_\alpha$ around 121\,nm, with the additional coherent 
contribution of the entire Lyman series \cite[e.g.,][]{St05}. Hence the 
importance for developing numerical tools that are adequate for the 
modeling of these spectral series.

Recently, several instruments for detecting the polarization of the
solar spectrum in the UV have been developed or deployed. 
The Chromospheric Lyman-Alpha Spectro Polarimeter (CLASP; \citealt{Ka12}) 
rocket experiment successfully measured the scattering polarization in 
\ion{H}{1} Ly$_\alpha$ and its variation along the solar radius
\citep{Ka17}. The results of this experiment confirmed important predictions 
from theoretical modeling of the Hanle effect in this line, but also opened 
new questions about the structure and magnetic topology of the upper 
solar chromosphere \citep{Ka17}. Motivated by the CLASP success, and by
ongoing efforts in the modeling of the \ion{Mg}{2} h--k doublet in the solar 
transition region spectrum at 280\,nm \citep{BT12,Al16,dPA16}, the CLASP-2 
mission \citep{Na16} was proposed in order to measure the polarization 
produced by the joint action of scattering processes and the Hanle and 
Zeeman effects in these lines.

\begin{figure}[t!]
\centering
\includegraphics[height=3.25in]{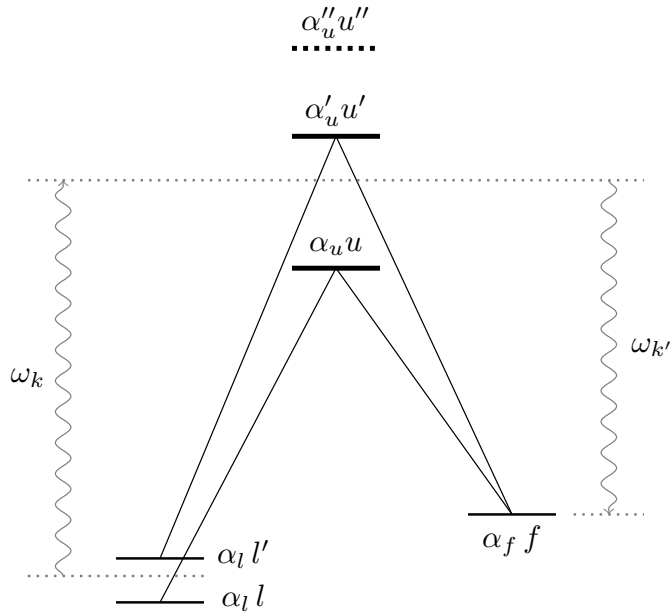}
\caption{\label{fig:model}
The ``multiple $\Lambda$'' atomic model, with lower terms $\alpha_l$ 
and $\alpha_f$, and upper terms $(\alpha_u,\alpha_u',\alpha_u'',\ldots)$. 
The illumination in the continuum can excite a virtual state 
between two contiguous configurations $\alpha_u$ and $\alpha_u'$, 
inducing coherence effects between them.}
\end{figure}

Other spectral structures of the solar UV spectrum offer additional 
insights in the energetics of the solar chromosphere and transition
region, such as the \ion{Fe}{2} UV multiplets between 230 and 260\,nm
\citep{AA89,Ju92}. These transitions originate from significantly 
more complex
atomic structures than \ion{H}{1} Ly$_\alpha$ and \ion{Mg}{2} h--k, and
the modeling of the expected polarization signatures from these
multiplets poses new challenges. One of them is to understand the
physics of how the \ion{Fe}{2} multiplets encompassed in this
spectral region interact with each other to shape the overall spectral
profile of the scattering polarization.

In this work, we rely on a recently proposed theoretical framework for
the description of partially coherent scattering of polarized radiation
\citep{Ca14,CM16a}, and extend it to the treatment of spectral series.
%
In Section~\ref{sec:theory} we develop the extension of this formalism,
and in Section~\ref{sec:application} we apply our results to the
modeling of the \ion{Fe}{2} UV multiplets 1--3 \cite[following the
classification of][]{Mo52}. In the conclusive section, we discuss our
findings, and provide a general physical explanation for the results,
with a demonstration fully presented in the Appendix.

\section{Radiation scattering in the multiple-$\Lambda$ atom}
\label{sec:theory}

We recall the expression of the polarized emissivity for a 
$\Lambda$-type atom, taking into account partially coherent scattering 
\citep{CM16a}, 
%
\begin{eqnarray} \label{eq:2emiss.1}
\varepsilon\apx{2}_i(\omega_{k'},\bm{\hat k}')
&=&\frac{4}{3}\frac{e_0^4}{\hbar^2 c^4}\,{\cal N}\omega_{k'}^4
	\sum_{ll'}\rho_{ll'}\sum_{uu'f}
	\sum_{qq'}\sum_{pp'}(-1)^{q'+p'}\,
	(r_q)_{ul}(r_{q'})^\ast_{u'l'}
	(r_p)_{u'f}(r_{p'})^\ast_{uf} \nonumber \\
&&\times
	\sum_{KQ}\sum_{K'Q'}\sqrt{(2K+1)(2K'+1)}\,
	\thrj{1}{1}{K}{-q}{q'}{-Q}
	\thrj{1}{1}{K'}{-p}{p'}{-Q'}\,
	T^{K'}_{Q'}(i,\bm{\hat k}') \nonumber \\
&&\times \int_0^\infty d\omega_k
	\left(
	\Psi_{u'l',ful}^{-k,+k'-k} + \bar\Psi_{ul,fu'l'}^{-k,+k'-k}
	\right)
	J^K_Q(\omega_k)\;,\qquad (i=0,1,2,3)
\end{eqnarray}
where $i=0,1,2,3$ indicates the four Stokes parameters $(I,Q,U,V)$.
The profiles $\Psi$ describe the frequency redistribution of the
incident radiation in the atomic rest of frame, and were
defined by \cite{Ca14}.
The transformation of the coherent emissivity
(\ref{eq:2emiss.1}) to the observer's frame of reference is easily 
accomplished via the formal substitution
\begin{equation}
\Psi_{u'l',ful}^{-k,+k'-k} + \bar\Psi_{ul,fu'l'}^{-k,+k'-k}
\to\mathrm{i}(\Omega_{u'}^*-\Omega_u)^{-1}
	R(\Omega_u,\Omega_{u'};
	\Omega_l,\Omega_{l'},\Omega_{f};
	\omega_k,\omega_{k'},\Theta)\;,
\end{equation}
where $R$
is the appropriate redistribution
function in the ``laboratory'' frame of rest for the problem at hand, 
$\Theta$ being the angle between the directions of the incident and 
emergent photons, and $\Omega_a=\omega_a-\mathrm{i}\epsilon_a$ the 
complex frequency of the atomic state $a$ with level width $\epsilon_a$. 
For the examples illustrated in 
Section~\ref{sec:application}, we employed the laboratory
frame redistribution function for the three-term atom of the $\Lambda$
type \cite[cf.][and Figure \ref{fig:model}]{CM16b}.
For the definition of all the other physical quantities in
equation~(\ref{eq:2emiss.1}),
the reader should refer to the paper of \cite{Ca14}.

In this section we extend equation~(\ref{eq:2emiss.1}) to 
the model of a multiple-$\Lambda$ atom (Figure~\ref{fig:model})
in the presence of a magnetic field. 
For simplicity, we derive the formalism for an atom without 
hyperfine structure, but the extension of the results to account 
for hyperfine structure is straightforward \citep{Ca14,CM16a}.

We indicate with $\alpha_l$ and $\alpha_f$ the electronic configurations 
of the initial and final terms of the $\Lambda$ transition, and with
$(\alpha_u,\alpha_u',\alpha_u'',\ldots)$ the set of intermediate upper terms
(Figure~\ref{fig:model}). We further assume the direction of the magnetic 
field as the quantization axis ($z$-axis). Then the atomic states involved 
in equation~(\ref{eq:2emiss.1}) are of the form
\begin{eqnarray} \label{eq:state.def}
l&\equiv&(\alpha_l \mu_l M_l)\;,\quad
l'\equiv(\alpha_l \mu_l' M_l')\;, \nonumber \\
u&\equiv&(\alpha_u \mu_u M_u)\;,\quad
u'\equiv(\alpha_u' \mu_u' M_u')\;, \\
f&\equiv&(\alpha_f \mu_f M_f)\;, \nonumber
\end{eqnarray}
where $M$ is the projection of the total angular momentum $\bm{J}=\bm{L}+\bm{S}$ on
the $z$-axis. The index $\mu$ spans the eigenspace of the atomic Hamiltonian
associated with a given value of $M$ \emph{and} term configuration 
$\alpha$. Hence, we assume a magnetic field regime such that 
configuration mixing induced by the magnetic Hamiltonian is negligible
\cite[see, e.g.,][Table~2, for an estimate of such condition in the case of the hydrogen
atom]{CL93}. We then can write
\begin{displaymath}
\ket{\alpha \mu M}=\sum_{J} C^{J}_\mu(\alpha M)\,
	\ket{\alpha JM}\;,
\end{displaymath}
where the (real) projection coefficients $C_\mu^J$ of the eigenstate
$\ket{\alpha \mu M}$ on the basis of the zero-field atomic states 
$\ket{\alpha JM}$ satisfy the orthogonality conditions
\begin{equation} \label{eq:diag.prop}
\sum_{J} C^{J}_\mu(\alpha M)\,C^{J}_{\mu'}(\alpha M)
	=\delta_{\mu\mu'}\;,\qquad
\sum_\mu C^{J}_\mu(\alpha M)\,C^{J'}_\mu(\alpha M)
	=\delta_{JJ'}\;.
\end{equation}

The density matrix element for the lower state, $\rho_{ll'}$, can be
written in terms of the irreducible spherical tensor components of the
statistical operator,
\begin{eqnarray} \label{eq:rho}
\rho_{ll'}
&\equiv&\bra{\alpha_l \mu_l M_l}\rho\ket{\alpha_l \mu_l' M_l'}  
	\nonumber \\
\noalign{\vspace{6pt}}
&=&\sum_{J_l}\sum_{J_l'}
	C^{J_l}_{\mu_l}(\alpha_l M_l)\,
	C^{J_l'}_{\mu_l'}(\alpha_l M_l') 
	\sum_{K_l Q_l} (-1)^{J_l-M_l}\sqrt{2K_l+1}\,
	\thrj{J_l}{J_l'}{K_l}{M_l}{-M_l'}{-Q_l}\,
	{}^{\alpha_l}\!\rho^{K_l}_{Q_l}(J_l,J_l')\;,
\end{eqnarray}
while use of the Wigner-Eckart theorem and its corollaries \citep[e.g.,][]{BS93}
gives the following expression for the dipole matrix element,
\begin{eqnarray} \label{eq:rdip1}
(r_q)_{ab}
&\equiv&\bra{\alpha_a \mu_a M_a}r_q\ket{\alpha_b \mu_b M_b} 
	\nonumber \\
\noalign{\vspace{6pt}}
&=&\sum_{J_a J_b}
	C^{J_a}_{\mu_a}(\alpha_a M_a)\,
	C^{J_b}_{\mu_b}(\alpha_b M_b)\,
	(-1)^{J_a-M_a}
	\sqrt{2J_a+1} 
	\thrj{J_a}{J_b}{1}{-M_a}{M_b}{q}
	\redmat{\alpha_a J_a}{\bm{r}}{\alpha_b J_b}\;.
\end{eqnarray}

To further proceed, we will assume that the multiple-$\Lambda$ atomic model is 
adequately described within the $LS$-coupling scheme, so that
$\alpha\equiv(\beta LS)$, where $\beta$ identifies the electronic 
configuration of a particular $LS$ term of the atom. We then can write, 
additionally,
\begin{eqnarray} \label{eq:rdip2}
\redmat{\alpha_a J_a}{\bm{r}}{\alpha_b J_b}
&\equiv&
	\redmat{\beta_a L_a S J_a}{\bm{r}}{\beta_b L_b S J_b}\;.
	\nonumber \\
&=&(-1)^{1+L_a+S+J_b} \sqrt{(2L_a+1)(2J_b+1)}\;
	\sixj{J_a}{J_b}{1}{L_b}{L_a}{S}\,
	\mathfrak{r}_{ab}\;,
\end{eqnarray}
where we introduced the (generally complex) reduced matrix elements of 
the dipole operator between orbital configurations
\begin{equation} \label{eq:strength}
\mathfrak{r}_{ab}=\redmat{\beta_a L_a}{\bm{r}}{\beta_b L_b}\;.
\end{equation}

Using equations~(\ref{eq:rho}) through (\ref{eq:strength}), 
equation~(\ref{eq:2emiss.1}) finally becomes
\begin{eqnarray} \label{eq:RT.J0}
\varepsilon\apx{2}_i(\omega_{k'},\bm{\hat k}')
&=&\frac{4}{3}\frac{e_0^4}{\hbar^2 c^4}\,{\cal N}\omega_{k'}^4
	\Pi_{L_l L_f} \sum_{\beta_u L_u}\sum_{\beta_u' L_u'} 
	\left(\mathfrak{r}_{ul}\,\mathfrak{r}^\ast_{uf}\right)\!
	\left(\mathfrak{r}^\ast_{u'l}\,\mathfrak{r}_{u'f}\right)
	\nonumber \\
&&\kern -1.5cm\times
	\sum_{J_u J_u' J_u'' J_u'''}
	\sum_{J_l J_l' J_f J_f'}
	\Pi_{J_u J_u' J_u'' J_u'''}\,
	\Pi_{J_l J_l' J_f J_f'}\,
	\sixj{J_u}{J_l}{1}{L_l}{L_u}{S}
	\sixj{J_u'}{J_l'}{1}{L_l}{L_u'}{S}
	\sixj{J_u''}{J_f}{1}{L_{l}}{L_u}{S}
	\sixj{J_u'''}{J_f'}{1}{L_{l}}{L_u'}{S} \nonumber \\
&&\kern -1.5cm \times
	\sum_{\mu_u M_u}\sum_{\mu_u' M_u'}\sum_{\mu_f M_f}
	C^{J_u}_{\mu_u}(M_u)\,
	C^{J_u''}_{\mu_u}(M_u)\,
	C^{J_u'}_{\mu_u'}(M_u')\,
	C^{J_u'''}_{\mu_u'}(M_u')\,
	C^{J_f}_{\mu_f}(M_f)\,
	C^{J_f'}_{\mu_f}(M_f) \nonumber \\
\noalign{\allowbreak}
&&\kern -1.5cm \times 
	\sum_{\bar J_l \bar J_l'}
	\sum_{\mu_l M_l}\sum_{\mu_l' M_l'}
	C^{J_l}_{\mu_l}(M_l)\,
	C^{\bar J_l}_{\mu_l}(M_l)\,
	C^{J_l'}_{\mu_l'}(M_l')\,
	C^{\bar J_l'}_{\mu_l'}(M_l') \nonumber \\
&&\kern -1.5cm \times
	\sum_{KQ}\sum_{K'Q'}\sum_{K_l Q_l}
	\sum_{qq'} \sum_{pp'}
	(-1)^{\bar J_l-M_l+q'+p'}
	\thrj{1}{1}{K}{-q}{q'}{-Q}
	\thrj{1}{1}{K'}{-p}{p'}{-Q'}
	\thrj{\bar J_l}{\bar J_l'}{K_l}{M_l}{-M_l'}{-Q_l}
	\nonumber \\
&&\mathop{\times}
	\thrj{J_u}{J_l}{1}{-M_u}{M_l}{q}
	\thrj{J_u'}{J_l'}{1}{-M_u'}{M_l'}{q'}
	\thrj{J_u'''}{J_f'}{1}{-M_u'}{M_f}{p}
	\thrj{J_u''}{J_f}{1}{-M_u}{M_f}{p'}
	\nonumber \\
&&\mathop{\times}
	\Pi_{KK'K_l}\,
	T^{K'}_{Q'}(i,\bm{\hat k}')\,
	\rho^{K_l}_{Q_l}(\bar J_l,\bar J_l') 
	\nonumber \\
&&\mathop{\times}
	\int_0^\infty d\omega_k
	\left(
	\Psi_{u'l',ful}^{-k,+k'-k} + \bar\Psi_{ul,fu'l'}^{-k,+k'-k}
	\right)
	J^K_Q(\omega_k)\;,\qquad (i=0,1,2,3)
\end{eqnarray}
where for simplicity of notation we omitted the term information 
from the density matrix element and from the argument of the 
projection coefficients $C^{J}_\mu$, and also adopted the shorthand notation
\begin{equation}
\Pi_{ab\ldots}\equiv\sqrt{(2a+1)(2b+1)\cdots}\;.
\end{equation}
%

The reduced matrix elements $\mathfrak{r}_{ab}$ must be evaluated for
each atomic species. In the case of complex atoms, this often requires
a numerical modeling of the atomic structure (e.g.,
Hartree-Fock, Dirac-Hartree-Fock, etc.) in order to determine the atomic
state wavefunctions. However, in the case of
hydrogenic atoms, those matrix elements are purely real, and can be
calculated with the aid of Gordon's formula \cite[][see also
\citealt{CL93}]{BS57}. This allows, for example, to use the emissivity 
(\ref{eq:RT.J0}) to model the \ion{H}{1} Lyman+Balmer
contribution to the polarization of the solar continuum.

In this paper, instead, we focus on spectral series with a common lower term, 
i.e., $\alpha_f=\alpha_l$, so the product of four $\mathfrak{r}_{ab}$ elements
in equation~(\ref{eq:RT.J0}) reduces to the product
$|\mathfrak{r}_{ul}|^2|\mathfrak{r}_{u'l}|^2$.
This can conveniently be expressed in terms of the Einstein $B_{lu}$ coefficient 
for absorption from the lower to the upper 
term of the atom, since
\begin{equation} \label{eq:Blu}
B_{lu}=\frac{16\pi^3}{3}\frac{e_0^2}{\hbar^2 c}\,
	\frac{\Pi_{L_u}^2}{\Pi_{L_l}^2}\,
	|\mathfrak{r}_{ul}|^2\;.
\end{equation}
Substitution of this equation into Equation~(\ref{eq:RT.J0})
leads to the appearance of a product $B_{lu} B_{lu'}$. 
On the other hand, it is customary to make the Einstein coefficient for 
spontaneous emission,
\begin{equation} \label{eq:Aul}
A_{ul}
=\frac{4}{3}\frac{e_0^2}{\hbar c^3}\,
	\omega_{ul}^3\,
	|\mathfrak{r}_{ul}|^2\;,
\end{equation}
also explicitly appear in the expression for the 2nd-order emissivity
\cite[e.g.,][]{CM16a}. 
This can be accomplished using the well-known relation
\begin{equation} \label{eq:AtoB}
B_{lu}=4\pi^3\,\frac{c^2}{\hbar\omega_{ul}^3}\,
\frac{\Pi_{L_u}^2}{\Pi_{L_l}^2}\,A_{ul}\;.
\end{equation}
It then becomes possible to rewrite the product $B_{lu}\,B_{lu'}$ 
using one of the following equivalent forms (among others), which are
symmetric with respect to the exchange of $u$ and $u'$,
\begin{eqnarray} \label{eq:BtoA}
B_{lu}\,B_{lu'}
&=&
2\pi^3\,\frac{c^2}{\hbar}\,\frac{1}{\Pi_{L_l}^2} \biggl(
       \frac{\Pi_{L_u}^2}{\omega_{ul}^3}\,A_{ul} B_{lu'}
      +\frac{\Pi_{L_u'}^2}{\omega_{u'l}^3}\,A_{u'l} B_{lu} \biggr)
\nonumber \\
&=&
4\pi^3\,\frac{c^2}{\hbar}\,\frac{\Pi_{L_u L_u'}}{\Pi_{L_l}^2} \biggl(
       \frac{A_{ul}\,A_{u'l}\,B_{lu}\,B_{lu'}}{\omega_{ul}^3\,\omega_{u'l}^3}
\biggr)^{1/2}\;.
\end{eqnarray}
Here we adopt the second form, through which the 2nd-order
emissivity (\ref{eq:RT.J0}) becomes,
\begin{eqnarray} \label{eq:RT.J}
\varepsilon\apx{2}_i(\omega_{k'},\bm{\hat k}')
&=&\frac{3}{16\pi^3}\,{\cal N}\hbar\,\omega_{k'}^4\,
	\Pi_{L_l}^2 \sum_{\beta_u L_u}\sum_{\beta_u' L_u'} 
\Pi_{L_u L_u'}\biggl(
       \frac{A_{ul}\,A_{u'l}\,B_{lu}\,B_{lu'}}{\omega_{ul}^3\,\omega_{u'l}^3}
\biggr)^{1/2} \nonumber \\
&&\kern -1.5cm\times
	\sum_{J_u J_u' J_u'' J_u'''}
	\sum_{J_l J_l' J_l'' J_l'''}
	\Pi_{J_u J_u' J_u'' J_u'''}\,
	\Pi_{J_l J_l' J_l'' J_l'''}\,
	\sixj{J_u}{J_l}{1}{L_l}{L_u}{S}
	\sixj{J_u'}{J_l'}{1}{L_l}{L_u'}{S}
	\sixj{J_u''}{J_l''}{1}{L_{l}}{L_u}{S}
	\sixj{J_u'''}{J_l'''}{1}{L_{l}}{L_u'}{S} \nonumber \\
&&\kern -1.5cm \times
	\sum_{\mu_u M_u}\sum_{\mu_u' M_u'}\sum_{\mu_l'' M_l''}
	C^{J_u}_{\mu_u}(M_u)\,
	C^{J_u''}_{\mu_u}(M_u)\,
	C^{J_u'}_{\mu_u'}(M_u')\,
	C^{J_u'''}_{\mu_u'}(M_u')\,
	C^{J_l''}_{\mu_l''}(M_l'')\,
	C^{J_l'''}_{\mu_l''}(M_l'') \nonumber \\
&&\kern -1.5cm \times 
	\sum_{\bar J_l \bar J_l'}
	\sum_{\mu_l M_l}\sum_{\mu_l' M_l'}
	C^{J_l}_{\mu_l}(M_l)\,
	C^{\bar J_l}_{\mu_l}(M_l)\,
	C^{J_l'}_{\mu_l'}(M_l')\,
	C^{\bar J_l'}_{\mu_l'}(M_l') \nonumber \\
\noalign{\allowbreak}
&&\kern -1.5cm \times
	\sum_{KQ}\sum_{K'Q'}\sum_{K_l Q_l}
	\sum_{qq'} \sum_{pp'}
	(-1)^{\bar J_l-M_l+q'+p'}
	\thrj{1}{1}{K}{-q}{q'}{-Q}
	\thrj{1}{1}{K'}{-p}{p'}{-Q'}
	\thrj{\bar J_l}{\bar J_l'}{K_l}{M_l}{-M_l'}{-Q_l}
	\nonumber \\
&&\mathop{\times}
	\thrj{J_u}{J_l}{1}{-M_u}{M_l}{q}
	\thrj{J_u'}{J_l'}{1}{-M_u'}{M_l'}{q'}
	\thrj{J_u'''}{J_l'''}{1}{-M_u'}{M_l''}{p}
	\thrj{J_u''}{J_l''}{1}{-M_u}{M_l''}{p'}
	\nonumber \\
&&\mathop{\times}
	\Pi_{KK'K_l}\,
	T^{K'}_{Q'}(i,\bm{\hat k}')\,
	\rho^{K_l}_{Q_l}(\bar J_l,\bar J_l') 
	\nonumber \\
&&\mathop{\times}
	\int_0^\infty d\omega_k
	\left(
	\Psi_{u'l',l''ul}^{-k,+k'-k} + \bar\Psi_{ul,l''u'l'}^{-k,+k'-k}
	\right)
	J^K_Q(\omega_k)\;.\qquad (i=0,1,2,3)
\end{eqnarray}

In the following section, we apply equation~(\ref{eq:RT.J}) to the 
modeling of the polarized emissivity of spectral series with a 
common lower term $\beta_l L_l$.

\section{Scattering Polarization of the \ion{Fe}{2} UV Multiplets}
\label{sec:application}

An excellent illustration of the effects of $L$-term interference
on the scattering polarization in spectral series, predicted by the
2nd-order emissivity (\ref{eq:RT.J}), is provided by the first three 
UV multiplets of \ion{Fe}{2} \cite[following the 
classification of][]{Mo52}, which are visible in the solar spectrum
between 230 and 265\,nm.

For our modeling, we assume that the atomic system is illuminated by 
a collimated beam of light (i.e., with maximum anisotropy) without 
spectral structure, 
and we compute the Stokes vector scattered at $90^\circ$ from the 
incidence direction. 
%
For the sake of demonstration, we only show calculations in the 
non-magnetic case, assuming a collisionless plasma at a 
temperature of $T=5000$\,K.
Since we have developed our formalism under the assumption that the 
magnetic field strength is small enough not to induce configuration 
mixing in the atomic system, magnetic effects, when included, will 
manifest around the transition resonances. In particular, they will 
appear as polarization signatures of the Hanle effect in the
transition cores, and of $J$-$J'$ interference within each multiplet 
in the near wings. Collective
effects encompassing the entire spectral series may also arise because 
of the common lower term of the series, which can be polarized, and 
therefore subject to the Hanle effect. However, such effects will also 
only manifest around the resonance frequency of the various atomic 
transitions in the series. 

\begin{figure}[!t]
\centering
\includegraphics[height=5in]{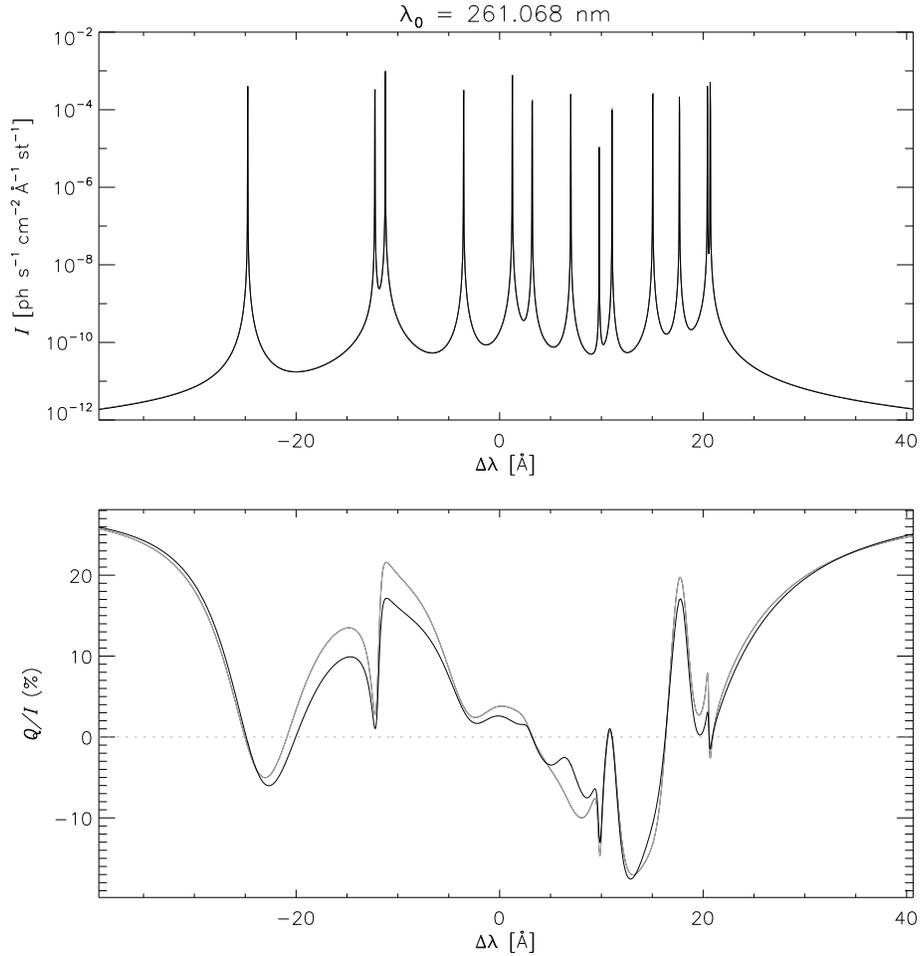}
\caption{\label{fig:prova_libro}
Intensity (top) and Stokes $Q/I$ polarization (bottom) in $90^\circ$ 
scattering and zero magnetic field, for the UV multiplet 1 of 
\ion{Fe}{2}. Gray curve: 
neglecting lower-level atomic polarization (l.l.p); black curve: taking
into account l.l.p. The bottom panel reproduces Figure~10.27 of
\cite{LL04}.}
\end{figure}

\begin{figure}[!t]
\centering
\includegraphics[height=5in]{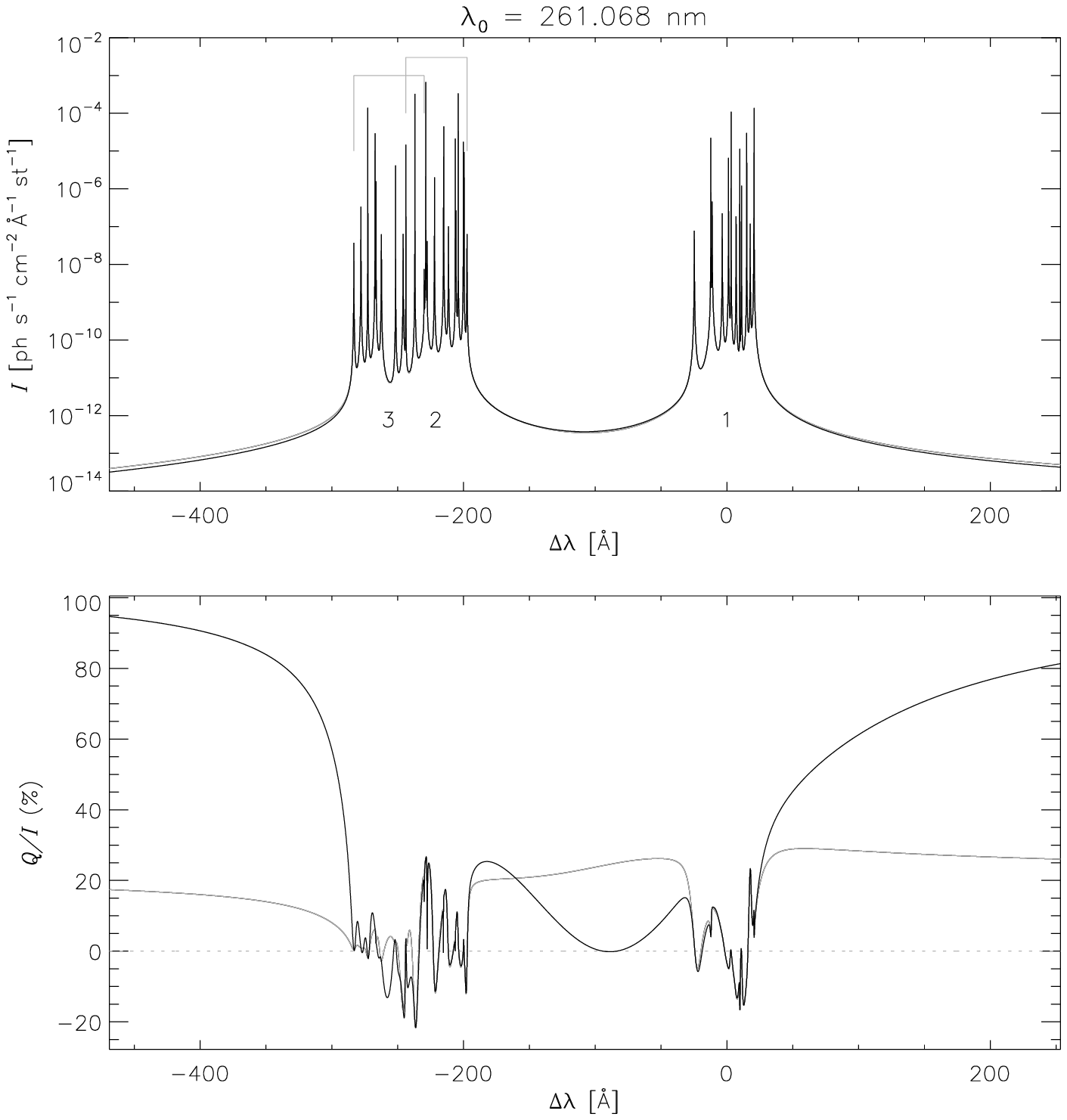}
\caption{\label{fig:fe_model}
Intensity (top) and Stokes $Q/I$ polarization (bottom) in $90^\circ$ 
scattering and zero magnetic field, for the set of UV multiplets 1,2,
and 3 ($\Delta L=L_u-L_l=0,+1,-1$, respectively) of \ion{Fe}{2}. We note that 
the component structures of multiplets 2 and 3 overlap, as indicated by
the bounding boxes drawn at the top. Gray curve: neglecting upper 
$L$-term interference; black curve: taking into account upper 
$L$-term interference.  We note the important depolarization effects 
around the center of gravity of the multiplet series, which are 
caused by the upper-term interference associated with 
the non-diagonal terms of the atomic density matrix. We also note 
the role of upper-term interference in causing the scattering 
polarization in the far wings of the spectral series to approach
the upper limit of 100\%.}
\end{figure}

Recently, \cite{Al16} and \citeauthor{dPA16} (\citeyear{dPA16}; see 
also \citealt{MS17}) 
have shown how PRD effects can combine with magneto-optical effects in 
an optically thick plasma to give rise to a significant $Q\to U$ 
``rotation'' of the scattering polarization in the near and far 
wings of deep resonance lines, even 
for magnetic strengths as small as a few gauss. Similar effects can be 
expected to manifest also for the model atom discussed in this paper. 
However, the necessary modeling effort involves the numerical solution 
of the full radiative transfer problem for polarized radiation in a 
realistic model atmosphere, which is beyond the scope of this paper. 

The configuration of the UV multiplet series of \ion{Fe}{2} presented
here is well described within the $LS$-coupling scheme, with $S=5/2$, and
a common lower term with $L_l=2$.
Figure~\ref{fig:prova_libro} shows the first UV multiplet at 261\,nm,
which corresponds to a $\Delta L=L_u-L_l=0$ transition, giving rise 
to 13
different $\Delta J$ transition components. The top panel shows the 
intensity amplitude of the scattered radiation in logarithmic scale, 
with all 13 transition components resolved. The lower panel shows 
the corresponding fractional $Q/I$ polarization. The gray profile in 
the lower panel models the case in which lower level polarization 
(l.l.p.) is neglected, whereas the black profile accounts for the
presence of l.l.p. The two models are also drawn in the intensity plot,
but they are essentially indistinguishable. The lower panel of
Figure~\ref{fig:prova_libro} reproduces Figure~10.27 of 
\cite{LL04}, and is presented here mainly as a 
test of our model.

Figure~\ref{fig:fe_model} shows all three multiplets of the series, with
multiplets 2 and 3 corresponding to the term transition $\Delta L=+1$ 
(14 components) and $\Delta L=-1$ (9 components), respectively. 
Multiplets 2 and 3 actually overlap in frequency, as shown in the
intensity plot of Figure~\ref{fig:fe_model}. These plots model
the polarization of the scattered light over the three multiplets,
assuming that the three corresponding term transitions are 
excited by the same 
collimated beam of unpolarized radiation, having a flat spectrum across 
the entire wavelength range of the series. The gray curve shows the 
emitted profile resulting from the incoherent addition of the 
individual contributions from the three multiplets. This is obtained by 
forcing the diagonality condition $L_u'=L_u$ in the 2nd-order 
emissivity (\ref{eq:RT.J}). The black profile instead takes into account
the quantum interference among the three upper terms of the series. 
In both models, the effects of atomic polarization in the ground term are
also fully accounted for. We note the striking difference in the
behavior of the polarization, both in the far wings and around the
center of gravity of the spectral group, when upper-term interference is
properly taken into account.


\section{Discussion and Conclusions}

\begin{figure}[t!]
\centering
\includegraphics[height=5in]{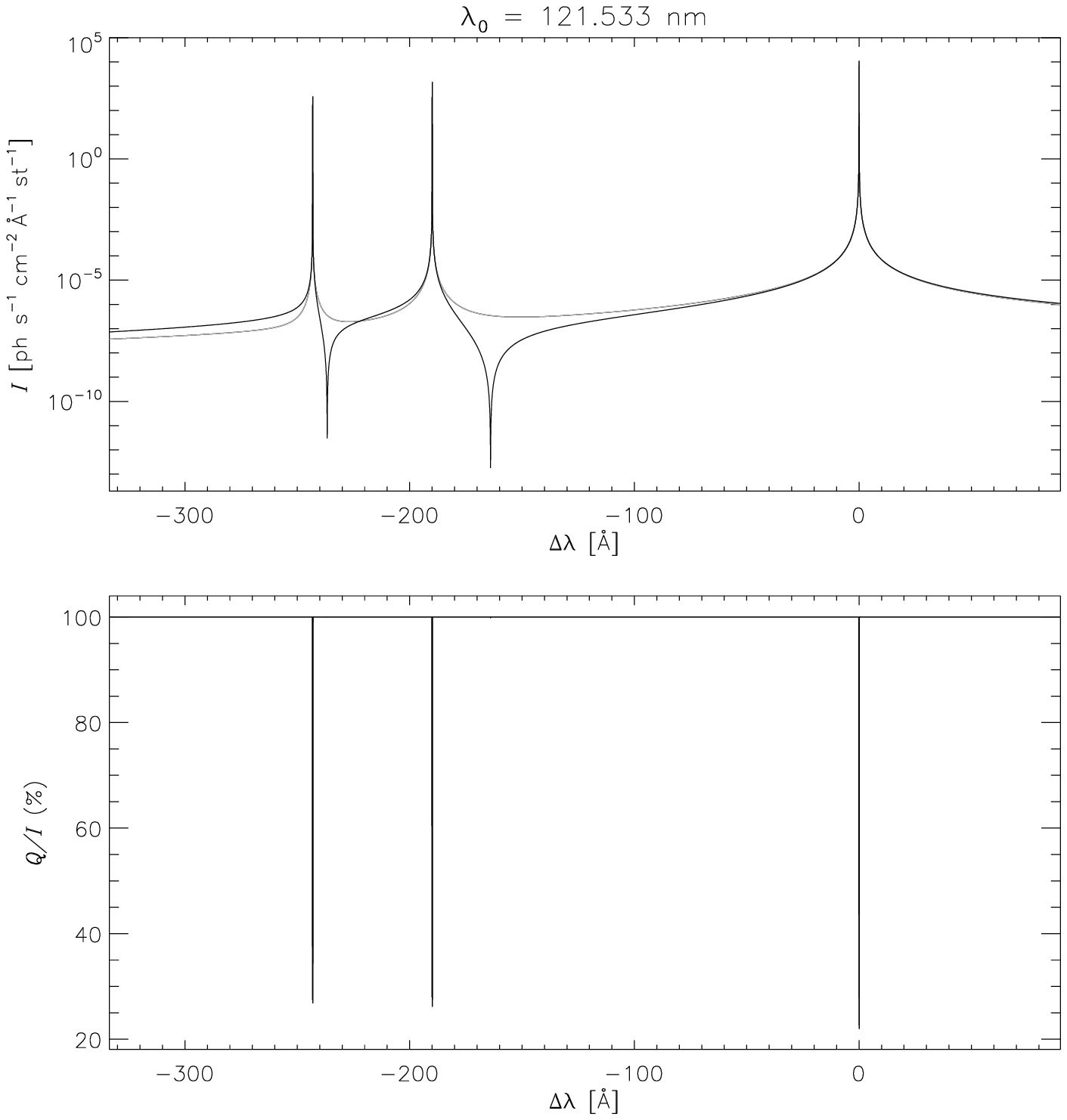}
\caption{\label{fig:lyman}
The first three lines of the \ion{H}{1} Lyman series.}
\end{figure}

\begin{figure}[t!]
\centering
\includegraphics[height=5in]{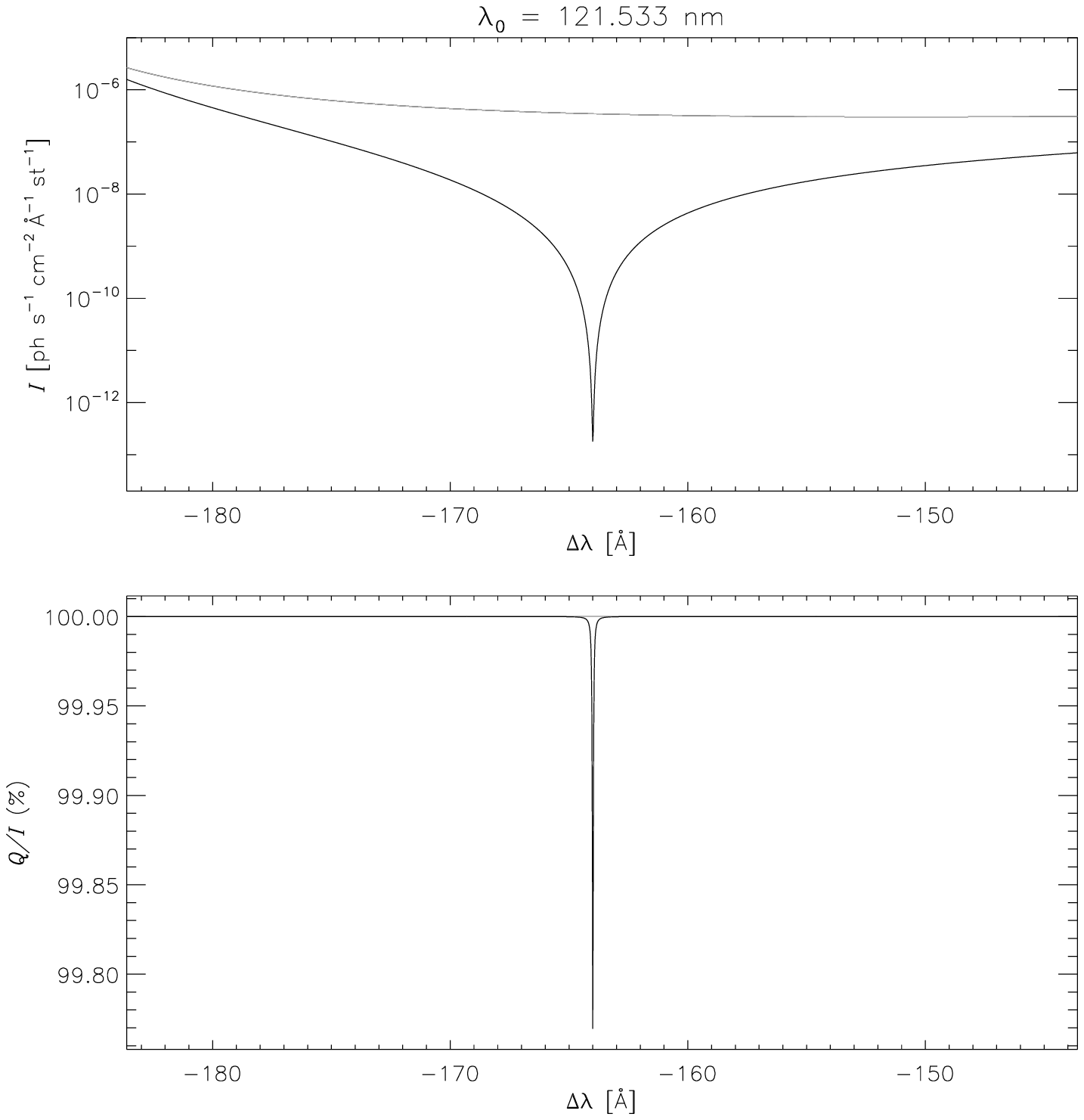}
\caption{\label{fig:lyman.det}
Spectral detail of the intensity dip structure in the red wing of 
\ion{H}{1} Ly$_\beta$.}
\end{figure}


The most notable feature of the modeled example of \ion{Fe}{2}
is the behavior of the 
$Q/I$ polarization in the far wings of the series in the presence of 
$L$-term interference. When moving toward the neighboring continuum, this 
polarization approaches the theoretical limit of 100\% expected for 
Rayleigh scattering---for the particular scattering model considered
here, where the atom is irradiated by a perfectly collimated beam of
unpolarized light. This is a manifestation of the so-called 
\emph{principle of spectroscopic stability} \cite[e.g.,][]{LL04}, 
according to which 
a complex system of atomic transitions between two fine-structured 
atomic terms must behave as a single transition between two simple 
terms with all the fine-structure details removed, in the 
experimental limit in which fine structure becomes unimportant. This 
is the case of the asymptotic behavior of the radiation scattered by
a multiplet at a distance from its center of gravity much larger 
than the frequency span of its fine structure. In order to satisfy 
this principle, the system of atomic transitions must be 
``complete'', in the sense that the considered set of fine-structure 
components must satisfy some kind of sum rule, once the spectral 
details are ignored. Closure of this sum rule requires that all 
possible interference terms between different levels be taken into 
account, a condition that is evidently satisfied by 
equation~(\ref{eq:RT.J0}). 

In the appendix, we provide a formal derivation of the asymptotic 
behavior of the Stokes profiles, in the particular case of
an unpolarized lower term, and determine the conditions 
under which the theoretical limit of 100\% polarization can be attained.
We find that a necessary closure condition is that 
$\Delta L$ attain all possible values allowed by the 
electric-dipole selection rule. When $L_l\ne 0$, these are
obviously $\Delta L=0,\pm1$, otherwise we only have $\Delta L=1$.

This last condition is verified in the case of the \ion{H}{1} Lyman
series, for \emph{each} individual transition in the series. This is 
clearly illustrated by Figure~\ref{fig:lyman}, where the first three Lyman
transitions of \ion{H}{1} are shown, both neglecting (gray curve) and
taking into account (black curve) $L$-term interference. In both cases,
the $Q/I$ polarization in the continuum between the lines attains the
theoretical maximum of 100\%. The intensity profile accounting for the
effects of $L$-term interference agrees qualitatively with the results
of \cite{St05}. Most notably, the model that takes into account the $L$-term
interference predicts the formation of intensity ``dips'' between lines
of the series. Figure~\ref{fig:lyman.det} shows the spectral details 
around the intensity dip in the right wing of Ly$_\beta$, from which we
see that these dips have a small depolarizing effect on the continuum.
For the modeling of Figures~\ref{fig:lyman} and
\ref{fig:lyman.det} we adopted the same model of collisionless plasma as in 
Section~\ref{sec:application}.

To conclude this section, we note that the presence of
the lower-term density matrix $\rho^{K_l}_{Q_l}(\bar J_l,\bar J_l')$
in the emissivity (\ref{eq:RT.J})
requires the numerical solution of the (first order) statistical
equilibrium (SE) problem for the polarized atom. 
In theory, in order for the model to be fully self-consistent, 
the SE problem should account for the same quantum interference effects 
among the upper terms of the spectral series that are included in 
the newly generalized form (\ref{eq:RT.J}) of the 2nd-order emissivity. 

For the numerical applications considered in this work, we did not 
generalize the SE problem in that sense. 
However, 
this does not affect the main results of this work, as the observed
qualitative behavior of the polarization of a spectral series is
reproduced even in the case of a naturally populated lower term 
(see Appendix).

\acknowledgments

We dedicate this work to the memory of our teacher, colleague, and
friend Egidio Landi Degl'Innocenti (1945--2017). He ``showed us the 
way''.
We thank P. Judge (HAO) and J. Trujillo Bueno (IAC, HAO) for helpful
discussion and comments on the manuscript.

\appendix

\begin{figure}[t!]
\centering
\includegraphics[height=3in]{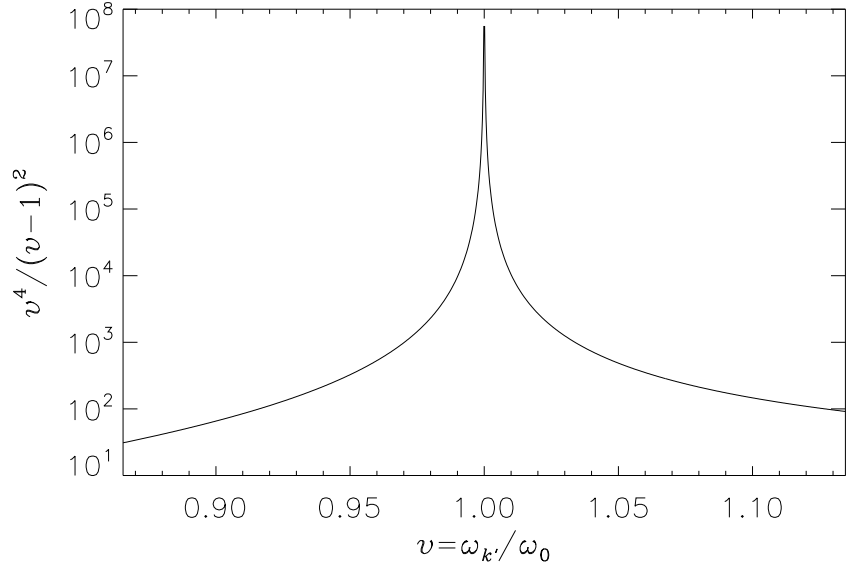}
\caption{\label{fig:rayleigh}
Plot showing the asymptotic behavior of the scattering profile
$\psi(\upsilon)=\upsilon^4/(\upsilon-1)^2$, where
$\upsilon=\omega_{k'}/\omega_0$. The adopted frequency range is comparable 
to the one adopted for Figure~\ref{fig:fe_model}.}
\end{figure}
\section{Asymptotic Limit of Scattering Polarization}

We consider the 2nd-order emissivity (\ref{eq:RT.J}) assuming that 
the lower term of the atomic system is unpolarized, with total
population $p_l$, so that 
\begin{equation} \label{eq:unpol}
\rho^{K_l}_{Q_l}(\bar J_l,\bar J_l')
=\delta_{K_l 0}\,\delta_{Q_l 0}\,\delta_{\bar J_l\bar
J_l'}\,\rho^0_0(\bar J_l)\;,\qquad
p_l=\sum_{\bar J_l} \Pi_{\bar J_l}\, \rho^0_0(\bar J_l)\;.
\end{equation}
We look at the asymptotic behavior of the scattered Stokes profiles,
i.e., at a frequency $\omega_{k'}$ such that the following
inequalities are both satisfied
\begin{equation} \label{eq:asympt.cond}
|\omega_{k'}-\omega_0|\gg|\omega_{uu'}|\ga\epsilon_{uu'}\;,
\qquad 
|\omega_{k'}-\omega_0|\gg|\omega_{ll'}|\;,
\end{equation}
where we indicated with $\omega_0$ the center of gravity of the system of
transitions. 

In the collisionless case, and under the usual assumption of a highly
diluted radiation field, $\epsilon_{l,l',l''}\to 0$ because of the 
implied infinite radiative lifetime of the lower levels.
%
Under these assumptions, and in the asymptotic limit defined by 
the conditions (\ref{eq:asympt.cond}), the redistribution function behaves 
like
\begin{equation} \label{eq:R2}
\Psi_{u'l',l''ul}^{-k,+k'-k} + \bar\Psi_{ul,l''u'l'}^{-k,+k'-k}
\sim 2\pi\,\frac{\delta(\omega_{k'}-\omega_k)}{(\omega_{k'}-\omega_0)^2}\;.
\end{equation}
(In order to see this one must consider the redistribution function (13) 
of \cite{Ca14}, noting that the conditions (\ref{eq:asympt.cond}) make the 
second term of that expression to vanish \emph{at least} as 
$(\omega_{k'}-\omega_0)^{-1}$, whereas the $\zeta$ functions in the first 
term force $\omega_k$ to also be in the same asymptotic regime as 
$\omega_{k'}$. This finally leads to the asymptotic limit (\ref{eq:R2}).)

Then, equation~(\ref{eq:RT.J}) becomes, after using the
orthogonality properties (\ref{eq:diag.prop}) and the integral norm of
the redistribution function \cite[cf.][equation~(15)]{Ca14},
\begin{eqnarray} \label{eq:A1}
\varepsilon\apx{2}_i(\omega_{k'},\bm{\hat k}')
&\sim&\frac{3}{8\pi^2}\,{\cal N}\hbar\,
	\frac{\omega_{k'}^4}{(\omega_{k'}-\omega_0)^2}\,
	\Pi_{L_l}^2 \sum_{\beta_u L_u}\sum_{\beta_u' L_u'} 
\Pi_{L_u L_u'} \biggl(
       \frac{A_{ul}\,A_{u'l}\,B_{lu}\,B_{lu'}}{\omega_{ul}^3\,\omega_{u'l}^3}
	\biggr)^{1/2} \\
&&\kern -1.5cm\times
	\sum_{J_u J_u'}
	\sum_{J_l J_l'}
	\Pi_{J_u J_u' J_l'}^2\,
	\Pi_{J_l}\,
	\rho^0_0(J_l)\,
	\sixj{J_u}{J_l}{1}{L_l}{L_u}{S}
	\sixj{J_u'}{J_l}{1}{L_l}{L_u'}{S}
	\sixj{J_u}{J_l'}{1}{L_l}{L_u}{S}
	\sixj{J_u'}{J_l'}{1}{L_l}{L_u'}{S} \nonumber \\
&&\kern -1.5cm \times
	\sum_{KQ}\sum_{K'Q'}
	\Pi_{KK'}\,
	J^K_Q(\omega_{k'})\,
	T^{K'}_{Q'}(i,\bm{\hat k}')
	\sum_{M_u M_u'} \sum_{M_l M_l'}
	\sum_{qq'} \sum_{pp'}
	(-1)^{q'+p'}
	\thrj{1}{1}{K}{-q}{q'}{-Q}
	\thrj{1}{1}{K'}{-p}{p'}{-Q'}
	\nonumber \\
&&\mathop{\times}
	\thrj{J_u}{J_l}{1}{-M_u}{M_l}{q}
	\thrj{J_u'}{J_l}{1}{-M_u'}{M_l}{q'}
	\thrj{J_u'}{J_l'}{1}{-M_u'}{M_l'}{p}
	\thrj{J_u}{J_l'}{1}{-M_u}{M_l'}{p'}\;,
	\qquad (i=0,1,2,3)
	\nonumber
\end{eqnarray}
We note that, despite the formal divergence of the ratio 
$\omega_{k'}^4/(\omega_{k'}-\omega_0)^2$ 
at infinity, the resulting line profile remains well behaved even 
in the far wings of the spectral series (Figure~\ref{fig:rayleigh}).

In considering the asymptotic behavior of the emissivity
(\ref{eq:A1}), the contraction of the full product of 3$j$-symbols 
becomes possible, and is easily accomplished by summing over 
$\{M_l,q,q'\}$, $\{M_l',p,p'\}$, and $\{M_u,M_u'\}$, in that order. 
We then obtain
\begin{eqnarray}
\varepsilon\apx{2}_i(\omega_{k'},\bm{\hat k}')
&\sim&\frac{3}{8\pi^2}\,{\cal N}\hbar\,
	\frac{\omega_{k'}^4}{(\omega_{k'}-\omega_0)^2}\,
	\Pi_{L_l}^2 \sum_{\beta_u L_u}\sum_{\beta_u' L_u'} 
\Pi_{L_u L_u'}\biggl(
       \frac{A_{ul}\,A_{u'l}\,B_{lu}\,B_{lu'}}{\omega_{ul}^3\,\omega_{u'l}^3}
	\biggr)^{1/2} \\
&&\kern -1.5cm\times
	\sum_{J_u J_u'}
	\sum_{J_l J_l'}
	(-1)^{J_l-J_l'}\,
	\Pi_{J_u J_u' J_l'}^2\,
	\Pi_{J_l}\,
	\rho^0_0(J_l)\,
	\sixj{J_u}{J_l}{1}{L_l}{L_u}{S}
	\sixj{J_u'}{J_l}{1}{L_l}{L_u'}{S}
	\sixj{J_u}{J_l'}{1}{L_l}{L_u}{S}
	\sixj{J_u'}{J_l'}{1}{L_l}{L_u'}{S} \nonumber \\
&&\kern -1.5cm \times
	\sum_{KQ} (-1)^Q\,
	\sixj{J_u}{J_u'}{K}{1}{1}{J_l}
	\sixj{J_u}{J_u'}{K}{1}{1}{J_l'}\,
	J^K_Q(\omega_{k'})\,
	T^{K}_{-Q}(i,\bm{\hat k}')\;.
	\qquad (i=0,1,2,3) \nonumber
\end{eqnarray}
The sums over $J_l'$ and $\{J_u,J_u'\}$ can also be performed, which 
leaves us simply with
\begin{eqnarray} \label{eq:form1}
\varepsilon\apx{2}_i(\omega_{k'},\bm{\hat k}')
&\sim&\frac{3}{8\pi^2}\,{\cal N}\hbar\,
	\frac{\omega_{k'}^4}{(\omega_{k'}-\omega_0)^2}\,p_l
	\sum_{KQ} (-1)^Q\,
	J^K_Q(\omega_{k'})\,
	T^{K}_{-Q}(i,\bm{\hat k}') \\
&&{}\times
	\sum_{\beta_u L_u}\sum_{\beta_u' L_u'} 
	\Pi_{L_u L_u'}\biggl(
        \frac{A_{ul}\,A_{u'l}\,B_{lu}\,B_{lu'}}%
	{\omega_{ul}^3\,\omega_{u'l}^3}
	\biggr)^{1/2}\,
	\sixj{L_u}{L_u'}{K}{1}{1}{L_l}^2
	\nonumber \\
\label{eq:last}
&=&\frac{3\pi}{2}\,{\cal N} c^2
	\frac{\omega_{k'}^4}{(\omega_{k'}-\omega_0)^2}\,
	\frac{p_l}{\Pi_{L_l}^2}
	\sum_{KQ} (-1)^Q\,
	J^K_Q(\omega_{k'})\,
	T^{K}_{-Q}(i,\bm{\hat k}') \\
&&\times
	\sum_{\beta_u L_u}\sum_{\beta_u' L_u'} 
	\Pi_{L_u L_u'}^2
	\frac{A_{ul}\,A_{u'l}}{\omega_{ul}^3\omega_{u'l}^3}\,
	\sixj{L_u}{L_u'}{K}{1}{1}{L_l}^2\;,
	\qquad (i=0,1,2,3) \nonumber
\end{eqnarray}
where we also used the definition of $p_l$ in
equation~(\ref{eq:unpol}), and in the second equivalence we 
used the 
relation (\ref{eq:AtoB}) in order to express the summation 
over the atomic configurations of the excited states exclusively 
in terms of the Einstein $A$-coefficients.
It is instructive to compare the form (\ref{eq:form1}) of the above
expression with equation~(10.148) of \cite{LL04},
which describes the asymptotic behavior of the emissivity for a two-term
atom, under the same hypotheses, and which can be used for modeling the
case of Figure~\ref{fig:prova_libro}.

We now consider the following two conditions: 1) the ratio
$A_{ul}/\omega_{ul}^3$ is independent of $u$,
and 2) the spectral series spans \emph{all and only} the values of $L_u$ 
that satisfy the triangular condition $\Delta L=0,\pm 1$, i.e., the
spectral series is \emph{complete}. When $L_l=0$, such as 
in the \ion{H}{1} Lyman series, the triangular condition becomes 
simply $L_u=1$, since $0\to0$ transitions are strictly
forbidden. Hence, \emph{each} Lyman transition is \emph{complete}
in itself. 

When both of the above conditions are met, either of the sums over $L_u$ 
or $L_u'$ in equation~(\ref{eq:last}) gives rise to the orthogonality 
(or closure) relation for the 6$j$ symbols \cite[e.g.,][]{BS93}, adding
up to 1, whereas the remaining sum evaluates to $3\,\Pi_l^2$. In this 
case, the asymptotic $Q/I$ polarization is simply given by the ratio
\begin{equation} \label{eq:asympt}
\frac{Q}{I}\sim
\frac{\sum_{KQ} (-1)^Q\,J^K_Q(\omega_{k'})\,T^{K}_{-Q}(1,\bm{\hat k}')}%
     {\sum_{KQ} (-1)^Q\,J^K_Q(\omega_{k'})\,T^{K}_{-Q}(0,\bm{\hat k}')}\;.
\end{equation}
In particular, for $90^\circ$ scattering by an atom illuminated with a 
collimated beam of unpolarized radiation, it can easily be shown that
the above ratio equals 1.

In the case of the \ion{Fe}{2} model of Figure~\ref{fig:fe_model}, the
invariance of $A_{ul}/\omega_{ul}^3$ across the spectral series is only 
coarsely satisfied, since that quantity is in the ratios 48:42:36 for the 
UV multiplets 1, 2, and 3, respectively. Yet, even under such loose 
condition,
the theoretical limit of $Q/I$ as calculated through equation~(\ref{eq:last})
still lies above 99\%.

The theoretical limit (\ref{eq:asympt}) breaks down when the spectral
series is not complete, in the sense specified above (e.g., in the
case of the single multiplet of Figure~\ref{fig:prova_libro}), or when
$L$-term quantum interference is neglected, which is the case shown by the
gray curve in Figure~\ref{fig:fe_model}. In fact, in such case the 
double summation over electronic configurations in equation~(\ref{eq:last})
takes the diagonal form
\begin{displaymath}
\sum_{\beta_u L_u}
	\Pi_{L_u}^4
	\sixj{L_u}{L_u}{K}{1}{1}{L_l}^2\;,
\end{displaymath}
which no longer corresponds to the closure relation for the 6$j$
symbols.

\end{document}